# Wide field of view crystal orientation mapping of layered materials

A. Orekhov[1,2,‡], D. Jannis[1,2,‡], N. Gauquelin[1,2], G. Guzzinati[1,2], A. Nalin Mehta[3,4],

S. Psilodimitrakopoulos[5], L. Mouchliadis[5], P. K. Sahoo[6], I. Paradisanos[6],

A.C. Ferrari[6], G. Kioseoglou[5,7], E. Stratakis[5,7,8], J. Verbeeck[1,2*]

[1] Electron Microscopy for Materials Science (EMAT) University of Antwerp 2020 Antwerp, Belgium

[2] NANOlab Center of Excellence, University of Antwerp, Belgium

[3] Imec, Kapeldreef 75, 3001, Leuven, Belgium

[4] KULeuven, Celestijnenlaan 200D, 3001, Leuven, Belgium

[5] Institute of Electronic Structure and Laser, Foundation for Research and Technology-Hellas, Heraklion Crete 71110, Greece

[6] Cambridge Graphene Centre, University of Cambridge, Cambridge CB3 0FA, UK

[7] Department of Materials Science and Technology, University of Crete, Heraklion Crete 71003, Greece

[8] Department of Physics, University of Crete, Heraklion Crete 71003, Greece

**ABSTRACT**

Layered materials (LMs) are at the centre of an ever increasing research effort due to their potential use in a variety of applications. The presence of imperfections, such as bi- or multilayer areas, holes, grain boundaries, isotropic and anisotropic deformations, *etc.* are detrimental for most (opto)electronic applications. Here, we present a set-up able to transform a conventional scanning electron microscope into a tool for structural analysis of a wide range of LMs. An hybrid pixel electron detector below the sample makes it possible to record two dimensional (2d) diffraction patterns for every probe position on the sample surface (2d), in transmission mode, thus performing a 2d+2d=4d STEM (scanning transmission electron microscopy) analysis. This offers a field of view up to 2 mm$^2$, while providing spatial resolution in the nm range, enabling the collection of statistical data on grain size, relative orientation angle, bilayer stacking, strain, *etc.* which can be mined through automated open-source data analysis software. We demonstrate this approach by analyzing a variety of LMs, such as mono- and multi-layer graphene, graphene oxide and MoS$_2$, showing the ability of this method to characterize them in the tens of nm to mm scale. This wide field of view range and the resulting statistical information are key for large scale applications of LMs.

**INTRODUCTION**

Graphene and related materials (GRMs) are at the centre of an ever increasing research area due to their set of unique properties[1]. Wafer scale growth is necessary for their integration in devices[1–4]. There is thus a need for fast (tens of minutes) and mm-scale (wafers scale) structural characterization. In single layer graphene (SLG) structural defects can significantly affect its properties[5,6]. E.g., grain boundaries and mismatch may affect the mechanical strength[7]. Since SLG is the thinnest monolayer (1L) that can be produced amongst all layered materials (LMs), we use it as a reference for the analysis of methods that provide structural and spatially resolved information on LMs. Raman spectroscopy is an ideal tool to non-destructively probe SLG's electronic and vibrational properties[8] and links these to structural information such as orientation, strain and number of layers, doping, defects[8]. Raman spectroscopy is noninvasive and can be used over large areas, in production plants[9]. Conventional light microscopy methods which rely on the oxidation of Cu underneath SLG, are used to detect the grain boundaries[10].

However, quantitative information on the orientation of grains is lacking[11]. Atomic scale study of GRMs is often performed with scanning probe microscopy (SPM)[12,13] or transmission electron microscopy (TEM)[14–16]. Both techniques resolve individual atoms, but suffer from small fields of view (<100x100 $nm^2$)[17]. In selected area electron diffraction (SAED), diffraction patterns are recorded in a TEM offering a slightly larger (200x200 $nm^2$ with the smallest 10μm selected area aperture) field of view[18–20]. Scanning electron microscopes (SEM) can achieve nm resolution[21], while being able to scan large (up to tens $mm^2$) fields of view[22]. In a conventional SEM, imaging is most commonly performed with backscattered[23] or secondary electrons[22], probing the sample surface[22]. SEM typically uses electron beam energies from 500 eV to 30 keV[22], considerably lower than the ~100-300 keV used in TEM[24]. For GRMs, the increased interaction strength at lower (<60 keV) energies results in more scattering[25], giving higher contrast (compared to TEM imaging) for materials such as SLG[25–27]. Compared to TEMs, SEMs are more common and easier to operate (requiring lower investment in both personnel and instrumentation), as no elaborate alignment procedure is required[28]. This makes it possible to fully automate the data acquisition process and makes SEM attractive for quality control on a production line. Indeed, "so-called" critical dimension SEM tools are installed in every wafer fab[29]. Furthermore, the larger working volume of the SEM vacuum chamber compared to TEMs offers space to install additional probes[30,31] and to operate on full (300 mm) wafer sized samples[32]. Environmental scanning electron microscopy (ESEM)[33] allows one to investigate materials in gaseous[34] or liquid[35] environments. Due to these advantages over TEMs, there is a growing interest in using SEM in transmission mode by inserting a detector behind an electron transparent sample to collect transmitted electrons[36–41]. Another emerging trend is the use of hybrid pixel detectors (semiconductor direct detection layer separated from the read out electronics[42]), offering advantages in the acquisition of diffraction patterns[43–45]. These cameras can detect single electrons with almost no thermal noise[46] and offer frame rates up to tens of kHz[47]. They are beam-hard (i.e. typical TEM electron energies do not destroy the detector)[46], and provide an extremely large dynamic range (e.g. intensity levels per pixels have 24 bit depth in Medipix3 based cameras)[47]. This allows rapid ~1000 frames per second (fps)[46] acquisition of multiple diffraction patterns without the need to mechanically block the central undiffracted beam (several orders of magnitude more intense than the diffracted ones[48]), unlike scintillator-based detectors[49].

Here, we present a wide field of view setup based on a SEM integrated with a hybrid pixel detector. The detector is placed in the far-field region below the sample and records diffraction patterns for each position of the electron probe. These patterns contain a wealth of structural information, while the spatial resolution is determined by the size of the electron probe. Such a setup is known as 4d-STEM [48–52], since we collect 2d diffraction patterns when scanning an electron probe over a 2d sample area. This allows us to perform structural analysis of GRMs over mm-scale fields of view, while retaining nm-scale resolution. We can extract grain size and orientation, detect twisted layers, bilayer stacking and unit cell deformation in minutes, providing valuable information for GRM monitoring at an industrial scale.

**RESULTS AND DISCUSSIONS**

**Experimental set up and samples**

We use a standard JEOL JSM-5510 SEM with a W electron source. We select this instrument type, since this very common, with an estimated 60000 installed similar instruments worldwide, enabling wide adoption of our proposed setup. Inside the vacuum chamber, samples are mounted on a Thorlabs 30 mm optical cage system, giving an adjustable sample to detector distance (camera length[24]) ranging from 5 to 50 mm, Fig.1. Diffraction pattern acquisition is performed using an Advacam®MiniPix detector with a 300 $\mu$m Si active layer[50]. This is a 256x256 hybrid pixel detector based on a Timepix 1 chip[51], integrated with its readout electronics in a USB stick format[52]. In combination with the fixed 55 $\mu$m pixel size of the detector, this results in a minimum resolvable angle~1.1mrad. For an electron beam of 20 keV, this corresponds to a pixel size with a resolution in reciprocal space of 1 pixel every 1/7 nm$^{-1}$. The scan coils are controlled using a custom scan engine, which allows synchronizing the detector readout with the probe positioning[53,54].

The working distance is ~18 mm, resulting in a convergence angle ~4.8±1.1 mrad (at 20 keV). The dead time, i.e. the time after each frame during which the detector cannot record another frame, is 20ms, which is a lower bound on the dwell time[50]. The small form factor (88x19x10 mm[50]) and low cost (<€5000 at 2020 rates[50]) of this device offsets the drawback of the large dead time, while paving the way to >100 times faster commercially available detectors, combining the same detector ASIC (Application-Specific Integrated Circuit) with a higher performance readout system reaching a few thousand fps[46]. Even though the scan speed is slow (dwell time~25ms), the acquired data has high signal to noise ratio and low background. The SEM is mechanically stable at this (up to 9000x) magnification and no knock-on damage is induced during acquisition, since the incoming electron energy is far below the threshold for SLG knock-on damage (~80keV)[55]. During the experiments, no structural damage is observed, with no noticeable effect on diffraction patterns. However, contamination forms around the probe position, alleviated by heating for 10 mins at 120 ºC prior to inserting the sample in the SEM. The main bottleneck is the limited framerate ~45 fps. A camera upgrade would allow the recording 1024x1024 scans within minutes[46], significantly lowering beam damage and electron beam induced deposition of contamination.

The detector efficiency (i.e. the ratio of detected electrons and total number of incident electrons[56]) approaches 80% for 20keV electrons on a 300$\mu$m Si active layer[57] (~20% are backscattered[23]) and the construction is intrinsically radiation hard[58]. This means that oversaturating some pixels with the primary beam does not affect the others, eliminating the need for a mechanical beam stopper[24]. As the beam scans the sample, the diffraction pattern is translated as well, making the use of a beam stopper difficult. As the total scan area is limited to 2 mm$^2$ field of view, this movement on the relatively large camera (14x14mm$^2$) is dealt with in software.

Ref. [59] used a fluorescence screen underneath the sample offering good dynamic range (12 bit) for the camera (albeit nonlinear) and beam hardness. However, the detection efficiency was~2%, which is disadvantageous for beam sensitive materials, such as zeolites[60] and metal-organic frameworks (MOFs)[61]. Ref. [62] used a masked single pixel detector matching the geometry of the diffraction pattern to do orientation mapping of SLG in SEM. This has the advantage of very fast response ($\mu$s per scan position[62]), but cannot cope with more complicated changes in crystal structure, such as strain[63] and crystal tiltmeasure[64], and makes very poor use of the incoming electron dose (with <10% of first order of diffraction spots transmitted through the mask).

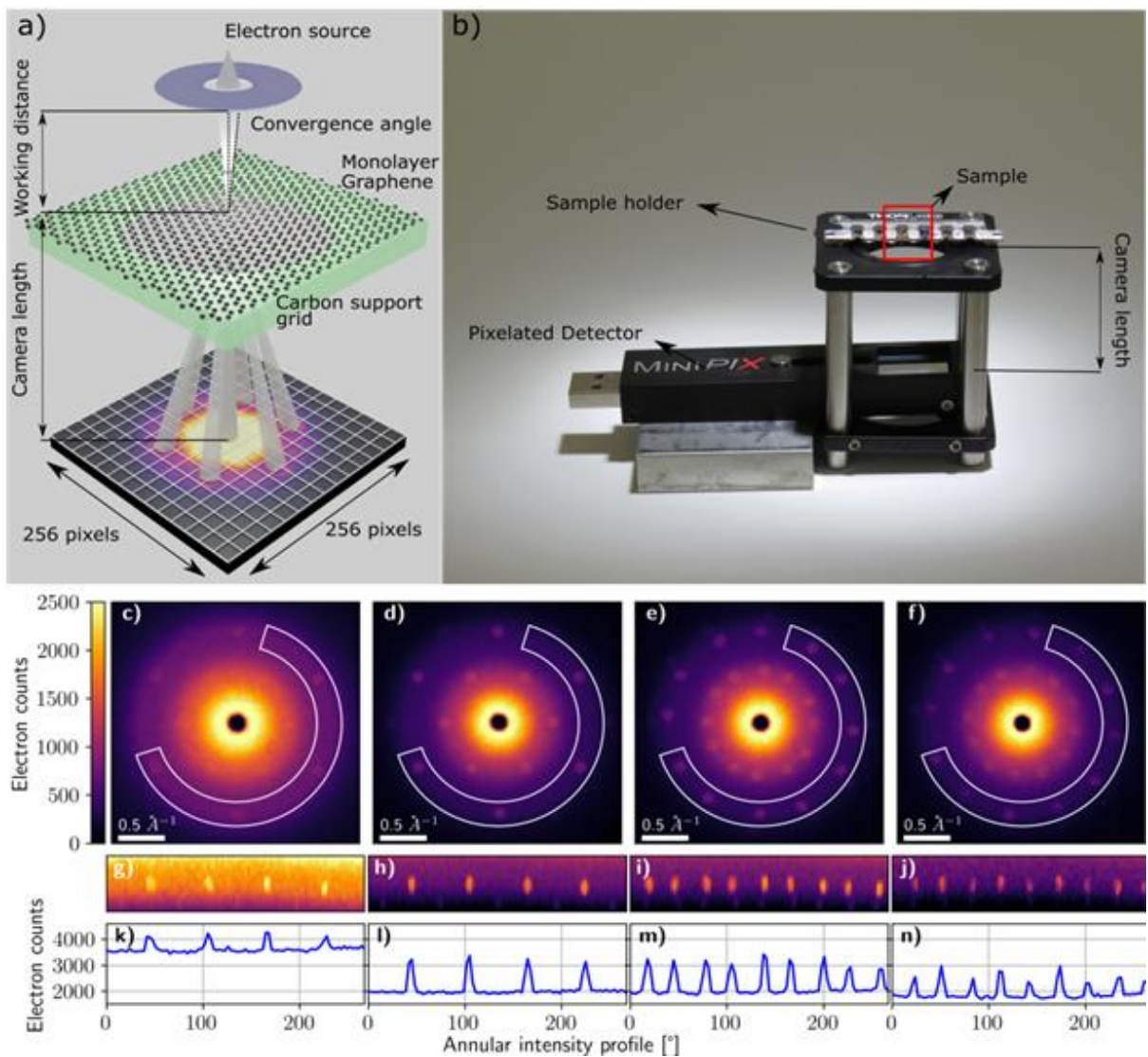

**Fig. 1. (a)** Schematic drawing of the experiment, consisting of a LM sample and a hybrid pixel detector at the bottom, inside an SEM sample chamber. **(b)** Picture of the setup. The sample mount allows camera length adjustment. **(c-f)** Examples of far-field electron diffraction patterns of SLG/BLG. (c) SLG on substrate, (d) free standing SLG, (e) BLG, (f) SLG grain boundary. The
7

central beam is oversaturated in terms of count-rate, therefore no intensity is observed. This does not affect neighboring image regions. **(g-j)** Azimuthal intensity profile of the second order spots for the corresponding diffraction pattern(above). **(k-n)** Radial integration over second order spots reveals differences in electron counts related to peak to background ratio and number of peaks.

We consider 4 types of representative LMs: SLG, Bilayer graphene (BLG), graphene oxide (GO) flakes, and mono (1L)- and bilayer (2L) $MoS_2$. Suspended SLG on TEM grids from Graphenea is used as representative commercial SLG[65]. Commercial GO is sourced from Sigma-Aldrich (763705-25ML). A GO and $H_2O$ is centrifuged and drop-cast onto a quantifoil TEM grid with amorphous carbon[66]. 1L-$MoS_2$ is grown by metal organic chemical vapor deposition (MOCVD) on sapphire[67], then transferred on a TEM grid via ultrasonic delamination, using polymethyl methacrylate (PMMA) as support layer[68]. 2L-$MoS_2$ flakes with 3R stacking are prepared on $SiO_2$/Si via chemical vapor deposition (CVD), and wet-transferred using PMMA followed by a controlled lift off process onto TEM grids[69–71], see Methods for details.

Figs.1c-f shows 4 diffraction patterns of SLG/BLG. The intensity of the central diffraction spot is suppressed, since the count rate of the incoming electrons (~2.5 x$10^6$/s on the central pixels) is too large for the detector to handle. Figs.1c,d show the SLG/BLG diffraction supported by a holey carbon membrane and over a freestanding hole. The pattern originating from the substrate supported area shows a larger background signal stemming from diffraction on the amorphous carbon substrate. In Figs.1e,f, twice as many diffraction spots are visible. The difference between these two panels is that in Fig.1e the intensity in every spot is comparable to the SLG diffraction pattern (Fig.1d), whereas in Fig.1f the intensity in each spot is lower. Hence, we interpret Fig.1e

as BLG with a twist angle~25° and Fig.1f as SLG where the probe is simultaneously illuminating two grains of different orientations, hence the intensity in the spots is shared over both patterns.

This shows the potential of the technique to derive information on number of layers (N), orientation, presence of amorphous material (contamination or support film)[72–74]. This paves the way to the quantitative investigation of sample cleanliness, e.g. important for SLG's use in electron microscopy as supporting film for the material under investigation[75,76], for gas sensors[77,78], and for liquid cells used during in situ TEM observations[79–81].

**Orientation mapping in SLG**

To demonstrate the effectiveness of the method, Fig. 2 plots a set of orientation maps of SLG with a field of view from mm to nm scale. 4 maps are acquired at different magnifications centered around the same location by decreasing the magnification from 9000x to 300x, which translate into scanned areas from 100 to 50000 $\mu m^2$. The orientation is determined at every probe position by analyzing each individual diffraction pattern, as discussed in Methods, and the results are presented in Figs.2a-d. Grain boundaries and average orientation per grain are identified by applying a 2d gradient operator on the orientation data, followed by a watershed algorithm of the Python library scikit-image[82] to segment the image. We define a grain boundary as given by a minimum of 2° difference in orientation in order to stay well above the noise level of the orientation determination step. From this segmented data, a rich variety of information is obtained, such as mean orientation, grain size, average grain shape, grain boundaries, etc. In Figs.2a-c the corresponding areas (inside a white rectangle) are identified, demonstrating negligible beam damage from the consecutive scans. The field of view of tens of thousands $\mu m^2$ could lead to automated scans of wafer size areas, while only spending time on the interesting regions, as long as a reasonable indicator for "interesting" can be formulated.

In order to achieve a mm-scale field of view, but have spatial resolution able to resolve $\mu$m size grains of a polycrystalline sample, a 1024x1024 probe position scan is performed at 250x magnification (Fig.2e). The number of different grains identified at this magnification is ~4200 over ~0.168mm². This gives access to unprecedented statistics on structural information, missing in the common tens $\mu$m2 field of view TEM investigations[24]. Fig.2f plots the grain size distribution. The average grain size is ~42$\mu$m and the 50th percentile is~22$\mu$m. The maximum detected grain size is~840$\mu$m² (white square in Fig.2e). As an example of mining this rich data, a correlation between grain size and orientation can be extracted by classifying these in 3 categories, corresponding to small~0.6-13$\mu$m, medium ~13-32$\mu$m, and large ~32-192$\mu$m grains. These intervals are chosen to show the evolution of different parameters with grain size. For large grains, we detect two preferential orientations, 30° apart from each other. Fig.2g plots the distribution of the 3 size classes. The solid lines are the result of fitting the distribution to the sum of two Gaussians and an offset. With increasing grain size, the orientations of the grains are progressively taking two preferential directions (22° and 52°) seen by the increasing amplitude of the Gaussian peaks compared to the constant background (see SI S4 for quantitative results).

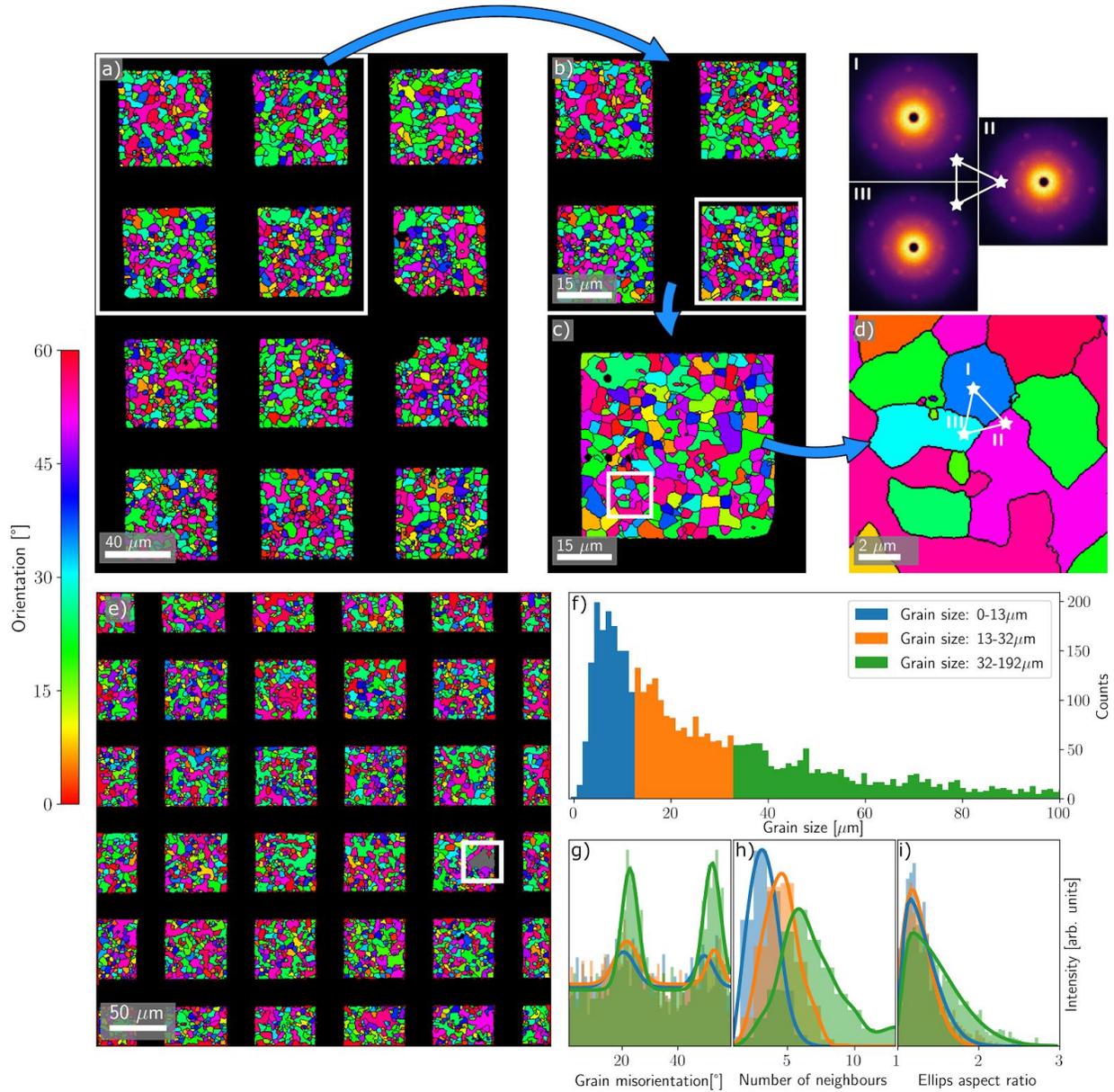

**Fig. 2.** Orientation mapping of a SLG over a wide range of fields of view: from mm-scale (250x magnification) to nm-scale (9000x). **(a-d)** Orientation maps at different magnifications of the same area of the sample, showing no noticeable beam damage. **(a)** 50000$\mu$m² field of view (300x magnification). **(b)** Zoom on the 4 squares at the top left of (a) scanned with 512x512 probe positions. **(c)** Zoom on the bottom right square of (b) using a 512x512 scan raster, corresponding to a field of view~3600 $\mu$m². **(d)** High magnification scan (9000x) with a raster of 256x256 probe

positions over~100 $\mu m^2$, corresponding to the area indicated with a blue square in panel (c). **(I-III)** 3 diffraction patterns from adjacent grains identified by the same indices in panel (d). **(e)** 1024x1024 orientation mapping of commercial SLG[9] over~0.168 $mm^2$ (250x magnification). The largest grain is indicated with a white square. **(f)** Grain size distribution calculated from the segmented dataset. **(g)** Correlation between grain size and orientation, indicating a preferential orientation for larger grains. The solid lines are the fitted results to the sum of two Gaussians and an offset. **(h)** Distribution of the number of neighbors as a function of grain size showing an increase in neighbors with increasing grain size. The solid line is a smoothed line through the data points. **(i)** Analysis of aspect ratio of grain shape as a function of grain size. The solid line is a skew normal fit through the data points (see SI S3 for details).

Another correlation is between grain size and number of adjacent grains, as more adjacent grains are expected around larger grains. Fig.2h plots the histograms corresponding to the different size categories. A correlation is observed, as the average number of surrounding grains is 3.29, 4.55, 7.39. The shape anisotropy of each grain can be determined by fitting its boundaries to an ellipse and taking the aspect ratio of its principal axes (see SI S4 for details). Fig.2i plots the aspect ratio for different grain sizes, where the distribution changes with grain size. The solid line is a skew-normal fit from which the mean value of the aspect ratio is~1.31, 1.39,1.46, indicating that larger grains are less spherical than smaller ones.

**$MoS_2$ orientation mapping over $mm^2$ range**

In order to demonstrate the applicability of the method to other LMs, such as transition metal dichalcogenides (TMDs), a MOCVD-MoS$_2$ sample is analyzed (Fig. 3). To achieve the largest field of view, without movement of the stage, we use the longest available working distance, and perform a 512x512 probe position scan on ~2.36 mm$^2$, with a dwell time ~40ms. The acceleration voltage, working distance, camera length and pixel size are 20 keV, 16 mm, 34 mm and 3 $\mu$m. This MOCVD-MoS$_2$ has N=1-3, with grain sizes in the nm range as can be seen in magnified virtual dark field image Fig.3d. Black contrast corresponds to lacey carbon support grid, while bright regions are several-layer grains of MoS$_2$. The large field of view scan cannot resolve the small grains since the probe step size is 3 $\mu$m, larger than the small grain size. Fig.3a shows the orientation map, indicating a single orientation (the positions of the diffraction spots remain the same across the scanned area), related to the epitaxial growth on sapphire giving the same crystal orientation with respect to the original substrate before exfoliation. Figs.3b,c plot MoS$_2$ diffraction patterns at two points ~1.3 mm apart, showing that the orientation is the same for both.

Another scan at 25000x magnification, where the field of view corresponds to 3 pixels in the large scan (50x), is acquired in order to check the MoS$_2$ microstructure. In Fig.3d, the intensity of the second order spots indicates the presence of multi layers (MLs). The orientation map of the high magnification scan in Fig.3e confirms that most 2 and 3L areas are epitaxially grown on top of the first layer. ~3% regions with different orientation are also present, in good agreement with 4d-STEM TEM measurements performed at ~1Mx magnification[83].

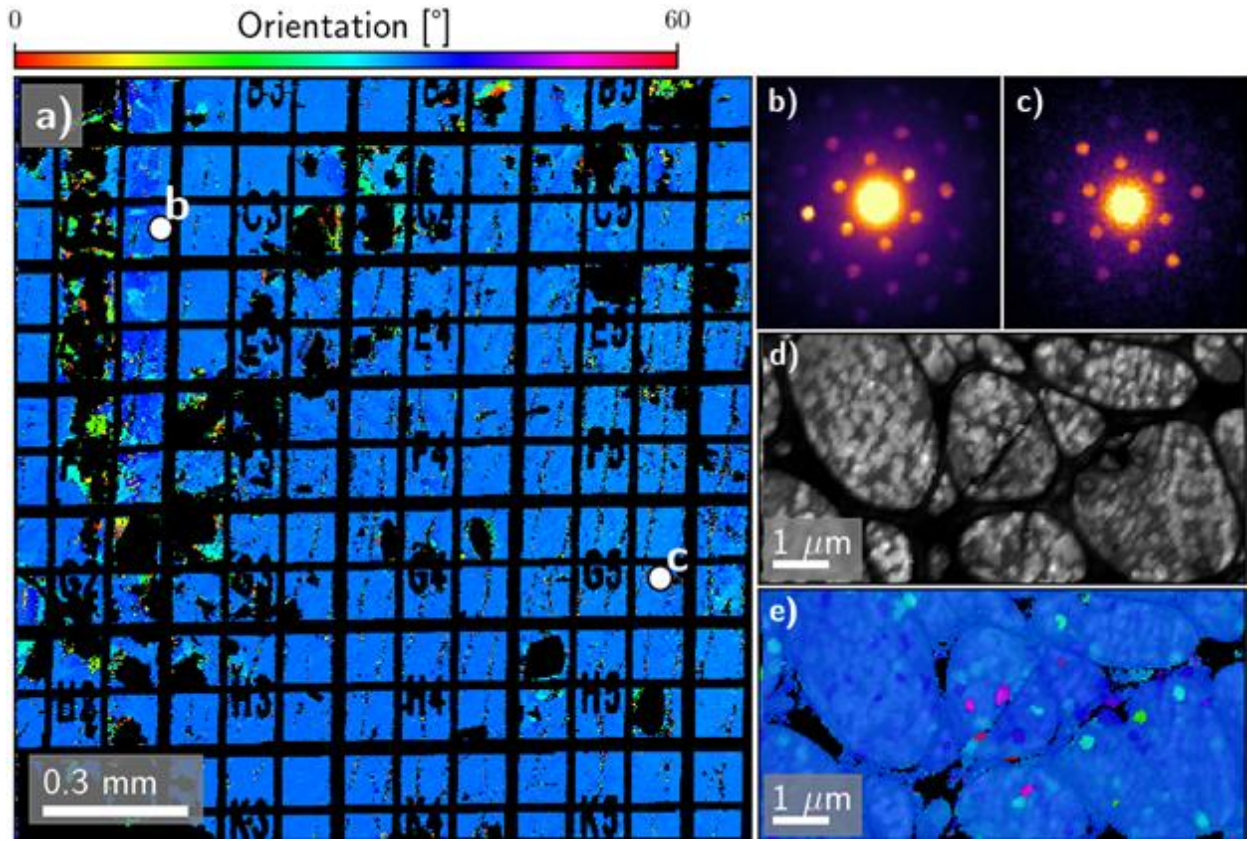

**Fig. 3. (a)** Orientation map of MoS$_2$ film, showing a single orientation (the positions of the diffraction spots remain the same across the scanned area) over a ~2.36 mm$^2$ field of view. **(b-c)** Diffraction patterns from positions indicated on (a). **(d)** The second order diffraction spot intensity mapping from scan at higher magnification (250x), where bright areas are 2 and 3L MoS$_2$ grains. **(e)** Orientation map (with overlapped VDF (d) for visual effect) showing that most 2 and 3L are epitaxially grown on top of the first layer.

**Correlation between orientation mapping and optical P-SHG microscopy**

The comparison of our approach with alternative optical methods, like second harmonic generation (SHG) imaging microscopy[84], provides a benchmark for the performance. We compare orientation mapping results with polarization-resolved SHG (P-SHG) microscopy[85,86] We use a CVD 2L-MoS$_2$ as test sample, transferred onto a carbon quantifoil grid and place this in both measurement setups, targeting the same sample areas. The 2L-MoS$_2$ sample has 3R stacking (see SI S6), with a broken inversion symmetry along the armchair direction[86]. This allows detectable SHG signals using e.g. photomultiplier tubes, since the second-order nonlinear optical susceptibility tensor is non-zero[87]. Mapping the armchair orientation gives us the same orientation information as 4d-STEM. A complete description of armchair orientation determination and instrumental parameters of P-SHG is in SI S6.

Fig.4a shows the Virtual Dark Field (VDF)[45] for a region with two MoS$_2$ flakes (marked with red and blue squares). The orientation map is in Fig.4b. The two regions have approximately the same orientation. By fitting the peaks with a Gaussian, the average difference in orientation between both grains is~0.2±0.7°.

The same area is analyzed by P-SHG (see SI S6 for the experimental setup), with the polarization of the incoming light undergoing a rotation from 0° to 360° in 2º steps. The integrated intensity map and the armchair orientation determined for every pixel are in Figs.4d,e. The average difference in the orientation is~1.7±1.9°, as seen from the histograms in Fig.4f. In our P-SHG experiments, the carbon quantifoil TEM grid, where the samples under examination are placed, is not compatible with the 1027nm, 90 fs, 76 MHz laser source used to excite the SHG signals and strong degradation of the TEM grid is observed. We thus use a low~1-2mW excitation power, resulting in very low SHG signals from MoS$_2$ (although MoS$_2$ has a large SHG response[87]) Thus, the peak broadening and 1.5° discrepancy in flake orientation between the two techniques could be attributed to the low signal to noise ratio of our SHG signals. Fig.4 shows that both methods have a similar domain of application, although significant differences exist. Nevertheless, the need for electron transparent samples in SEM is relieved for P-SHG (which does not need samples on a TEM grid), making it more suitable for inline quality control of LMs placed on thick (>100 nm) substrates. The spatial resolution is up to a factor 100 better for SEM, and both can deal with similar fields of view.

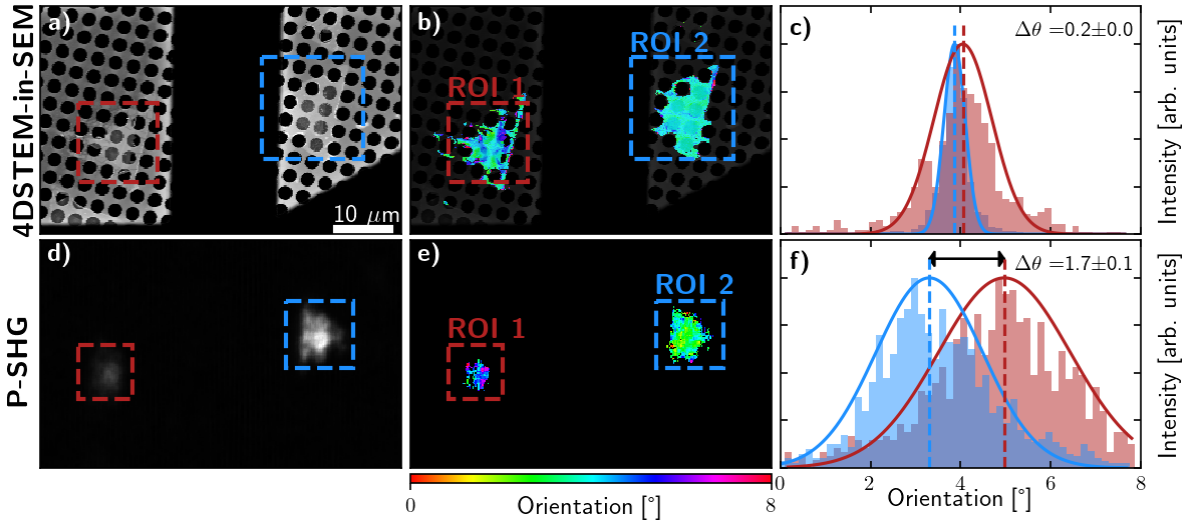

**Fig. 4.** Comparison of 4d-STEM in SEM and P-SHG data on CVD-MoS$_2$. **(a,d)** VDF signal and integration of P-SHG data for [0°–360°] with a 2° step. Regions of interest corresponding to two flakes are indicated. **(b, e)** Orientation map (with overlapped VDF image) based on 4d-STEM data and mapping of armchair orientations over a large (~2300 μm$^2$) sample area. **(c,f)** Histograms of two flakes orientation (colors correspond to selected areas in orientation map) from 4d-STEM and P-SHG, showing a qualitative match.

**Multilayer characterization**

We now consider the ability to probe ML systems by analyzing ML GO flakes. A scan of 512x1024 probe positions is performed over ~12800 $\mu m^2$. Direct visual identification of multiple GO layers is hardly accessible from the SE image, due to the large difference in sample/support thickness (~1/200), and lack of contrast, due to the comparably (to substrate thickness) minor change in thickness caused by the film stacking. 4d-STEM allows us to find interesting regions using offline diffraction data processing. After subtraction of the amorphous carbon background, VDF imaging allows us to visualize the regions where flakes are distributed, see SI S5.

A ML region is identified in the inset of Fig.5a. In order to determine N, a peak finding algorithm is applied in each diffraction pattern. Next, peaks at the radius of the second order spots are counted. 6 peaks indicate a 1L (AA and AB stacking excluded), while 12 peaks indicate 2L. An increase of intensity in the VDF (inset of Fig.5a) rules out grain boundaries, indicating 2L. Fig.5a shows the number of identified peaks and Fig.5c the relative orientation map. The minimum detectable twist angle depends on the convergence angle of the probe and the reciprocal pixel size of the detector, itself dependent on camera length, detector pixel size and accelerating voltage. When the 2L twist angle is zero, depending on the translation between the two 1L, these can be stacked AA and AB[88]. Simulation of electron diffraction patterns (see SI S2.) shows that N for AA, AB can be derived by taking the mean intensity of the second order spots, since these increase monotonically up to N~8 for 20 keV. For N>8, dynamical diffraction becomes dominant, redistributing the spot intensities nonlinearly.

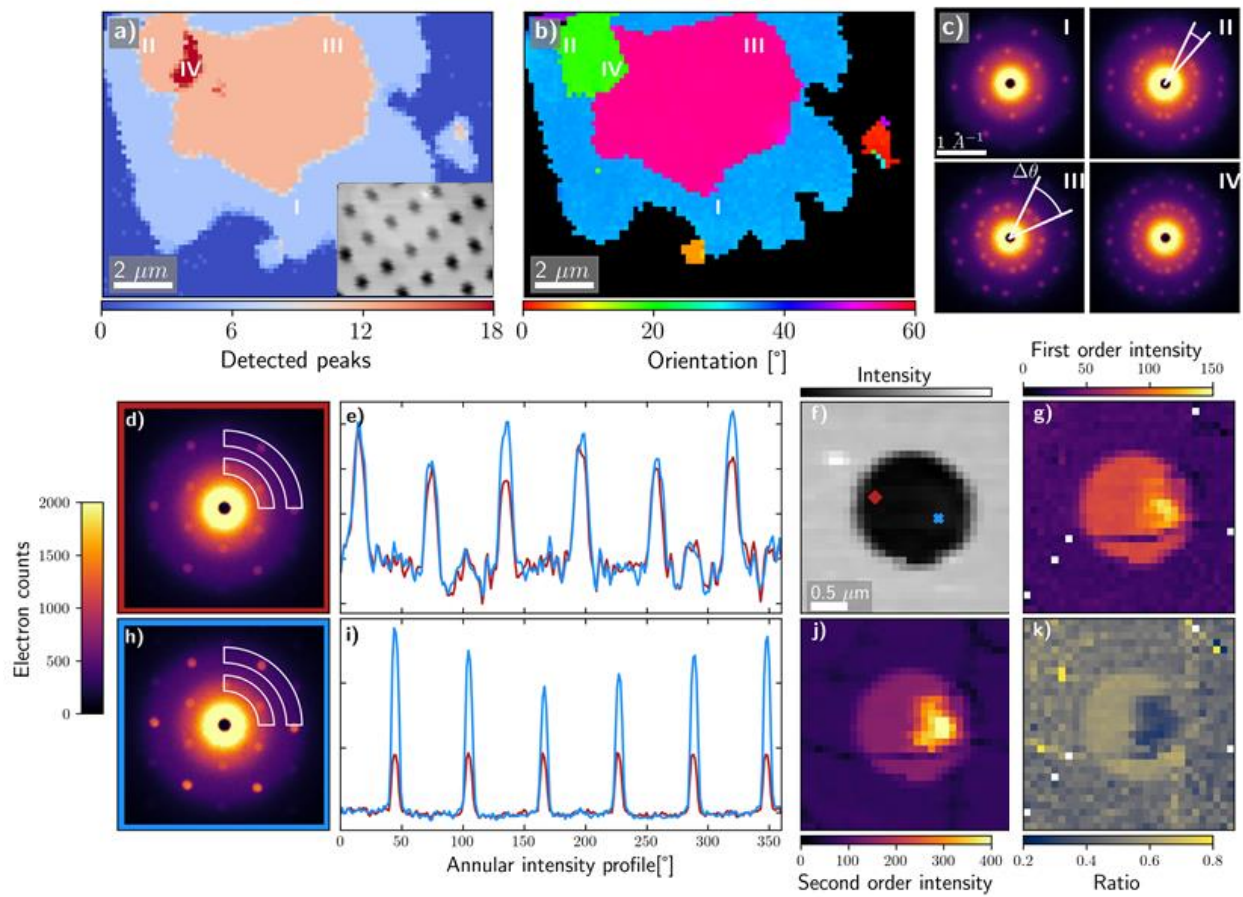

**Fig. 5. (a)** Number of peaks detected in the second order diffraction spots, indicating N, if the twist-angles is resolvable. For AB and AA stacking, N cannot be retrieved from the number of peaks, since no new spots arise with respect to 1L GO. The inset shows the VDF image. **(b)** Orientation map of GO sheets. **(c)** Diffraction patterns from regions where the twist angle is visible. The twist angle in 2L is indicated on the patterns. **(d,h)** Diffraction pattern of SLG and BLG, corresponding to probe positions indicated in (f). **(e,i)** BLG stacking analysis using the comparison of annular intensity profiles of the first (b) and second order (f) spots. AB stacking is identified by a ~2.5 times increase in the second order intensity, while no significant change of the first order intensity is observed. **(f)** VDF signal of the area mapped, as indicated in SI. (uppermost white arrow). **(g,j)** Average intensity map of first and second order spots. The area where the signals increase indicates MLs. **(k)** Identification of BLG stacking by taking the ratio between first and second order VDF images. (see Fig.S2 for comparison AA and AB stacking)

We now consider BLG. Diffraction patterns of SLG and BLG are shown in Figs.5d,h. For BLG an increase in the second order peak intensity is seen. The corresponding probe positions are indicated on Fig. 5f. The full scan of SLG is in SI S1, where the BLG region is indicated with a white arrow. To further investigate the BLG stacking, the annular intensity profiles of the first and second order peaks from SLG and BLG are plotted in Figs.5b,f. The intensity of the first order peaks does not show much variation, whereas the second order intensity increases significantly, indicating AB stacking for BLG. AA stacking would result in an increase of the same amount for the first order peaks[88], not seen here (Fig. S1a-b). Hence, the ratio between the first and second order intensities can reveal the type of stacking. This process can be applied to every diffraction pattern in the maps in Fig.5, where panels (d,g) are the average peak intensities of the first and second order peaks. In Fig.5g MLs can be identified as regions where an increase in signal is observed. To investigate stacking, the ratio between maps (d) and (g) is plotted in panel (h). A decrease of intensity in the BLG region is seen, identifying its AB stacking. For MLs other stacking sequences can occur, where the sequence identification becomes more elaborate[88].

**CONCLUSIONS**

By mounting an hybrid pixel detector in an existing SEM, we showed that 4d-STEM data can be obtained from various GRMs. Their few-layer thickness makes them highly suited for transmission electron diffraction, with no need of sample preparation even at beam energies of a few 10 keV, if the material is on an electron-transparent support, or is self-supported. The method provides a rich source of information on grain orientation and boundaries, twist-angle, stacking and number of layers. Scans of up to mm sized fields of view with a nm scale electron probe are possible. The geometrical projection of the diffraction pattern onto the detector provides a robust calibration of the scattering angles not influenced by lens drift and distortions. The data processing can be automated and runs over a full dataset unattended after an initial calibration setup. This requires~430 mins for acquisition and a few hours of data processing on a standard desktop PC for a 1024x1024 dataset. This can be decreased by using a detector offering a recording time reduction >100 times[46], with an even higher reduction expected for next generation Timepix and Medipix chips[89]. The data analysis time can be further decreased with software optimization and the use of more powerful computers. Our results have clear potential for use in industrial and research applications in metrology and quality control of GRMs and their heterostructures.

## METHODS

**Growth, transfer and characterization of MoS$_2$ flakes**

MoS$_2$ flakes are prepared via NaCl assisted CVD growth[90]. The precursor solution is prepared by mixing MoO$_3$ powder (90mg) and NaCl (10mg) in 5 mL of deionized (DI) water. The solution is sonicated for >10 min to prepare a uniform concentration of MoO$_3$ + NaCl (20 mg mL$^{-1}$). Few (~2-3) drops of MoO$_3$+NaCl are then spread on a cleaned Si+285nm SiO$_2$ substrate (6x12mm) and dried at 60 °C for 10 min to remove the water content. The MoO$_3$+NaCl films on SiO$_2$ is used as solid precursor for the growth of MoS$_2$. We use a fully automated temperature controlled two-zone CVD with a 60 mm diameter quartz tube and a pair of rollers to shift the heating zone (planarGROW-2S-MoS$_2$, planarTECH). The precursor facing upward is placed within a high pure (> 99.5%) alumina boat. 15 x 30 mm$^2$ SiO$_2$/Si substrates, pre-cleaned with acetone, isopropanol, and DI water followed by 1 min oxygen plasma treatment, are placed on the top of the alumina boat with the polished surface facing the precursor. The alumina boat is then placed in the middle of zone (II), while another boat containing 100 mg S powder is placed zone (I) in the upstream at a temperature of ~220 °C. Prior to growth, the CVD chamber is evacuated to 0.01 mTorr, and purged with high pure Ar (99.999%). Then, the pressure is kept ~100mTorr. The temperatures of the heating zones are raised to 100 and 300 °C, respectively, in 20 min with a constant flow of high pure Ar gas (200 sccm). Next, the temperature of the MoO$_3$ precursor is raised to 800 °C for the evaporation. Simultaneously, the sample temperature is raised to 220 °C for S evaporation. The growth is performed at 800 °C for 15 min under a constant flow of H$_2$ (10%) with Ar (200 sccm). When the temperature is below 700 °C, the growth process is terminated by sliding away the sample to cool down rapidly without altering the carrier gases. In this process, isolated 1 and 2L MoS$_2$ are obtained over an area of at least 1x1 cm$^2$.

MoS$_2$ flakes are then transferred from the SiO$_2$/Si substrates onto Au coated holey carbon TEM grids (hole diameter ~1.5 to 3μm) via wet transfer[69,70]. First, 2-3 drops of Poly(methyl methacrylate) (PMMA) solution are spin coated (~3000 rpm for 60s) on the substrate containing the as-synthesized MoS$_2$ flakes. The PMMA coated sample is dried at 65 °C for 10 min to improve adhesion between PMMA and MoS$_2$. Subsequently, the SiO$_2$/MoS$_2$/PMMA sample is immersed into 370mg KOH/10mL DI water. The typical time to remove MoS$_2$/PMMA from SiO$_2$ is~20-30mins at room temperature. The MoS$_2$/PMMA is then fished out using a clean glass slide and placed in DI water. Then, a 2x2mm$^2$ section is fished out on the Au coated holey carbon grid and dried in a desiccator overnight. The TEM grid with MoS$_2$/PMMA is then immersed in acetone, until it dries out, thus cleaning the PMMA layer and any organic residues.

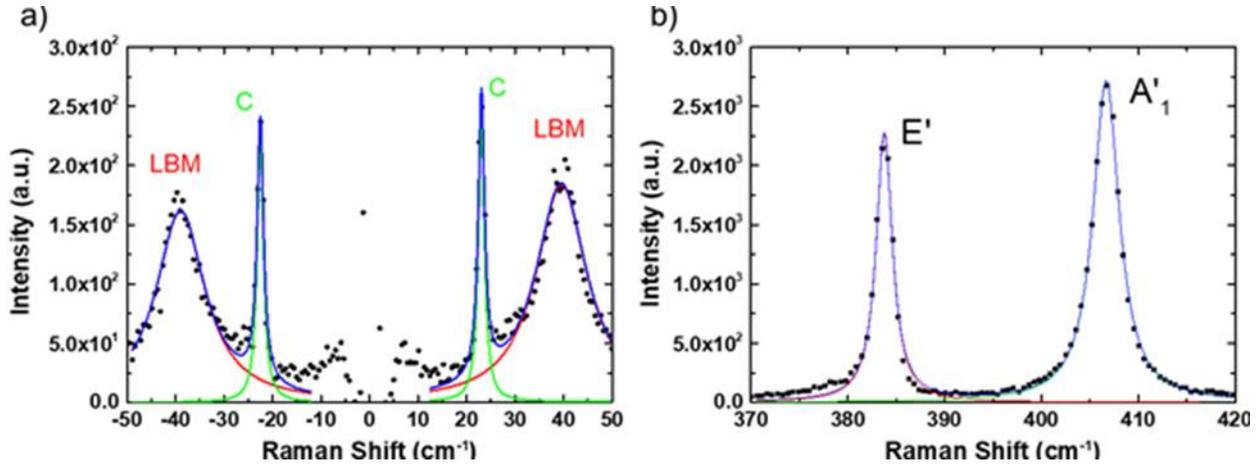

**Fig. 6. (a)** Raman spectrum of 3R-stacked CVD-2L-MoS$_2$. (a) LBM and C fitted with Lorentzians. The sum of the fit is shown blue. **(b)** E´ and A´$_1$ modes.

To evaluate N and stacking order of MoS$_2$ we perform Raman measurements at 295K, in a Horiba LabRam Evolution using 1800 l/mm grating and a 514nm excitation, with 60μW power to avoid heating effects. The spot size is ~1μm and the cut-off frequency is 5cm$^{-1}$. Fig.6a,b is the Raman spectrum of 2L CVD-MoS$_2$ after transfer on the carbon quantifoil grid. The layer-breathing (LBM) and shear (C) modes are observed in both Stokes and anti-Stokes, Fig.6a. Pos(C)~23cm$^{-1}$ is independent of the stacking order between 2H (180° twist angle) and 3R (0° twist angle), Pos(LBM) and the intensity ratio , verify the 3R stacking order[91,92]. Fig.6b shows in-plane (E´) and out-of-plane (A´$_1$) modes[91]. The difference in the Raman shift between E´ and A´$_1$~22.9cm$^{-1}$ further confirms N=2 [93].

**Determination structural information from electron diffraction patterns**

The acquisition of a diffraction pattern at every probe position retains all the information of the scattering in ($k_x$,$k_y$). Compared to conventional SEM, where secondary electrons are counted at every probe position[22], our setup generates a huge amount of data. In every probe position 256x256=65536 pixels are acquired. E.g., for a 1024x1024 scan, the data amounts to 80GB, which poses challenges for data storage and processing[94–96]. Therefore, there is a need for software which can handle these large data sets. We use the open-source python libraries Hyperspy[95] and Pixstem[94], developed to process 4d-STEM datasets with a desktop PC and automating the data processing while only having to tweak a minimal amount of parameters (experimental data and its treatment are available at [97]). Due to the large field of view, there is a movement of the diffraction patterns over the detector. This is corrected through image registration of each individual diffraction pattern by calculating the center of mass in every pattern and fitting this to a linear function. After this, any shape of virtual detectors (i.e. summing the intensity over a freely defined region) can be applied. This results in images representing specific contrasts. In Fig. 2a, the VDF[45] signal gives thickness (and density) contrast, revealing holes in the quantifoil film. In the middle of the image, a brighter rectangle results from a previous scan which induced contamination, increasing the local thickness.

More advanced data processing can be done to retrieve structural information by finding each diffraction peak position in every pattern and then fitting these points to the two reciprocal lattice vectors, $v_1$ and $v_2$, of the SLG hexagonal structure, resulting in a reciprocal matrix $G$[63]:

$$G = \begin{bmatrix} v_{1,x} & v_{2,x} \\ v_{1,y} & v_{2,y} \end{bmatrix}$$

Information on strain, orientation and shear is retrieved by obtaining the affine transformation matrix with respect to a reference reciprocal matrix $G_0$[63]:

$$G_0 = \begin{bmatrix} \langle ||v|| \rangle & \frac{1}{2}\langle ||v|| \rangle \\ 0 & \frac{\sqrt{3}}{2}\langle ||v|| \rangle \end{bmatrix}$$

The determination of $G_0$ is arbitrary[63]. Here we use the average norm of the fitted vectors as reference. G and $G_0$ can be linked with an affine transformation (A)[98]:

$$A = G_0 G^{-1}$$

The rotation ($\theta$), strain ($\varepsilon_{xx}$, $\varepsilon_{yy}$), shear ($\varepsilon_{xy}$) and unit cell expansion ($\varepsilon_{u.c.}$) are found by performing a polar decomposition on A[98]:

$$A = \begin{bmatrix} \cos\theta & -\sin\theta \\ \sin\theta & \cos\theta \end{bmatrix} \cdot \begin{bmatrix} c_{xx} & c_{xy} \\ c_{yx} & c_{yy} \end{bmatrix}$$

$$\varepsilon_{xx} = c_{xx} - 1 \qquad \varepsilon_{yy} = c_{yy} - 1 \qquad \varepsilon_{xy} = \frac{c_{xy} + c_{yx}}{2}$$

**ASSOCIATED CONTENT**

In Supporting Information presented additional information in sections: orientation, strain and diffraction spot intensity mapping; graphene multilayer simulations; ellipse fitting of the grains; fitting results grain orientation and aspect ratio histograms; identification of small particles/sheets distribution; determination of the armchair orientation in $MoS_2$ using P-SHG.

**AUTHOR INFORMATION**

**Corresponding Authors**

*E-mail: jo.verbeeck@uantwerpen.be

**Author Contributions**

A.O. and D.J contributed equally to this work. J.V. designed and managed the project. N.G. and G.G. helped in the development of an experimental setup for electron diffraction analysis and in data collection. A.N., P.S., I.P., A.F. performed the growth of samples and their characterization using Raman Spectroscopy. S.P., L.M., G.K., E.S. performed the characterization of samples by PSH-G method. All authors contributed to writing the paper and discussing the results.

**Notes**

The authors declare no competing financial interest.

**ACKNOWLEDGMENT**

Authors acknowledge to Prof. W. Vandervorst from KULeuven/IMEC for his input and stimulating discussions. Authors acknowledge funding from EU FLAG-ERA JTC 2017 GRAPH-EYE. J.V. and D.J. acknowledge funding from FWO G093417N ('Compressed sensing enabling low dose imaging in transmission electron microscopy') from the Flanders Research Fund. J.V. and N.G. acknowledge funding from EU No 823717 – ESTEEM3 and GOA "Solarpaint" of the University of Antwerp. GG acknowledges support from a senior postdoctoral fellowship grant from the Fonds Wetenschappelijk Onderzoek – Vlaanderen (FWO). We thank the Operational Program Competitiveness, Entrepreneurship and Innovation, under the call European R & T Cooperation-Grant Act of Hellenic Institutions that have participated in Joint Calls for Proposals of European Networks ERA NETS (National project code: GRAPH-EYE T8EPA2-00009 and European code: 26632, FLAGERA), EU Graphene Flagship, ERC Grants Hetero2D and GSYNCOR, EPSRC Grants EP/K01711X/1, EP/ K017144/1, EP/N010345/1, EP/L016087/1.

# Supplementary Information

# S1. ORIENTATION, STRAIN AND DIFFRACTION SPOT INTENSITY MAPPING

In Fig.S1b, the orientation of the two vectors obtained from A ranges from 0° to 60° due to the SLG six-fold symmetry. This reveals different grains, their boundaries and the average grain orientation. The underlying amorphous quantifoil grid with a thickness ~15nm has little effect on this procedure and grain orientation information is obtained over the whole field of view.

Information on strain and shear indicates the deformation of the unit cell. In Fig.S1c, the relative unit cell expansion ($\varepsilon_{u.c}$) is shown, in which a small difference is seen inside the hole. The sign of the expansion is positive indicating the increase of the unit cell in real space compared to the average unit cell. The intensity of the second order diffraction can be correlated with N when the twist angle is zero[1–3]. AA or AB stacking can be retrieved by taking the ratio of the first and second order intensities, and confirmed by diffraction pattern simulations (see S2.). The white arrows in Fig.2d indicate MLG patches, where the region located around the upper white arrow is investigated. If a twist is present, the diffraction spot intensity needs to be correlated with the number of peaks in order to distinguish between a grain boundary or BLG, Fig.1e-f. The anisotropy of the peak intensities indicates local tilts of SLG[4,5]. The peak intensity is calculated by first removing the background signal which stems from contamination, amorphous support regions and inelastic scattering. This background is obtained by taking the azimuthal median of every radius. After subtracting it from the raw signal, the pure diffraction contrast from SLG is extracted. This works due to the isotropic character of the background[6], combined to the sparsity of the diffraction signal where the sparsity originates from the periodicity of the SLG carbon atoms. Fig.2d is obtained by selecting the intensity of the most intense spot arising from the second order SLG diffraction. By comparing the intensity inside the hole or at the substrate, we see a higher intensity of the spots in the holes. Multiple scattering in the support film removes intensity from the diffraction spots. This gives an absorption loss ~75%. A decrease in signal is seen at the grain boundaries where, instead of one grain, two different grains are illuminated at one probe position, splitting the intensity over both orientations, Fig.1f. The high intensity lines across are due to wrinkles increasing local thickness and scattering[7,8].

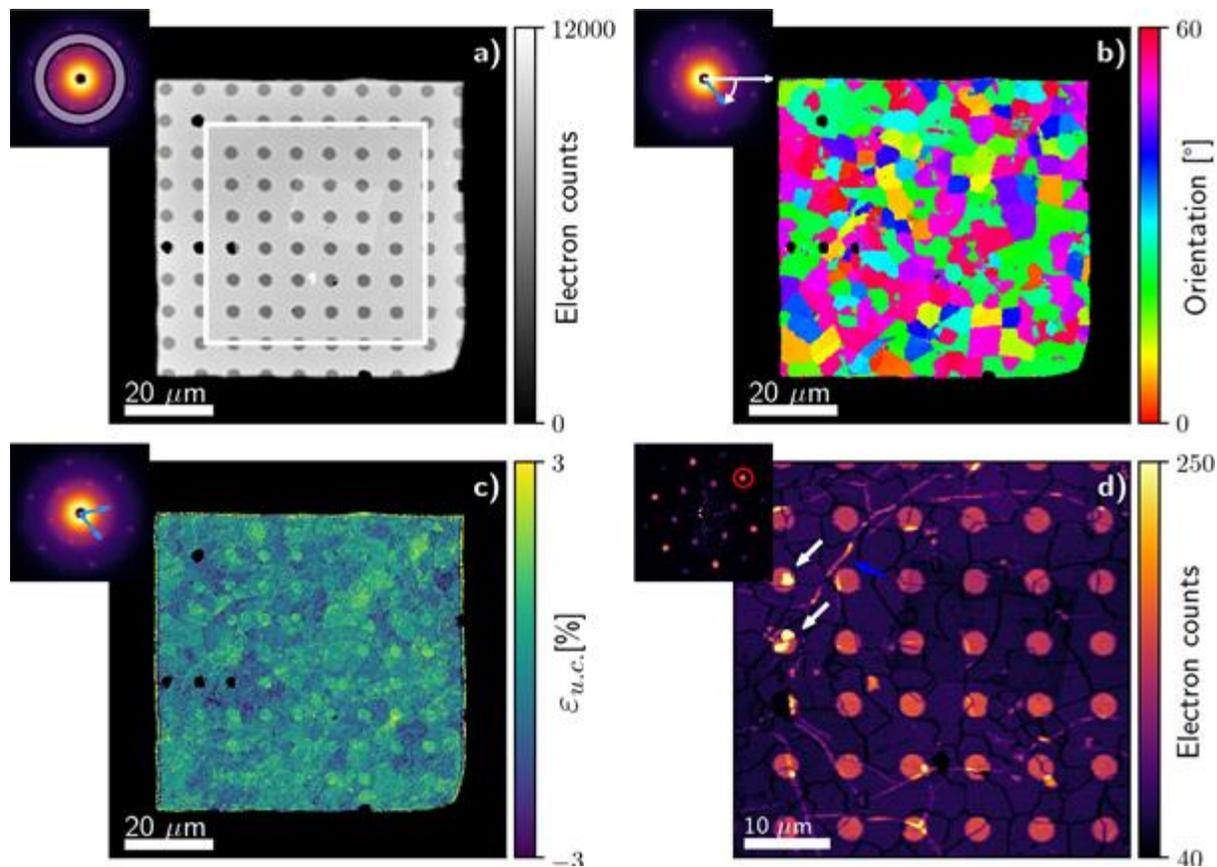

**Fig.S1. (a)** VDF images obtained from applying a virtually defined detector area, shown as inset. A scan of 512x512 probe positions with a dwell time ~40ms is obtained from the GO sample on a carbon quantifoil with a periodic set of holes. **(b)** Orientation mapping showing grain boundaries and mean orientation of each grain also in the areas where the support grid is present. **(c)** $\varepsilon_{u.c.}$ unit cell expansion compared to the average unit cell across the entire field of view. ~0.5% positive strain is seen in the holes, indicating an increase of unit cell area. **(d)** VDF image using the intensity

of the brightest second order spot showing a mixed contrast due to grain boundaries (lower intensity lines have the same contours from (b)), wrinkles (blue arrow) and multiple layers (white arrows). The smaller scan area indicated by a white rectangle in (a) is shown to increase visibility of the features. The inset is the background subtracted diffraction pattern.

## S2. GRAPHENE MULTILAYER SIMULATIONS

We now investigate the ability to detect MLG and AA or AB stacking by using simulations. The software used for electron diffraction simulations is Multem[9,10]. This allows one to include different approximations in the simulations, such as phase object[11] and multislice approximations[12]. Figs. S2a-b plots two BLG simulated diffraction patterns for (a) AA and (b) AB stacking. The convergence angle (3.5 mrad) and beam energy (20keV) are the same as in the experiments in Fig.1. The two patterns use the multislice approach[12]. The second order intensity is constant, while the first order diminishes for AB stacking. In Fig. S2c, the second order spots intensities are shown for different layers and AA and AB stacking. The difference between AA and AB stacking is minimal and increases monotonically up until N=8. For N>8 the phase object approximation breaks down because the sample becomes too thick. Hence, 20keV detects N=1-8.

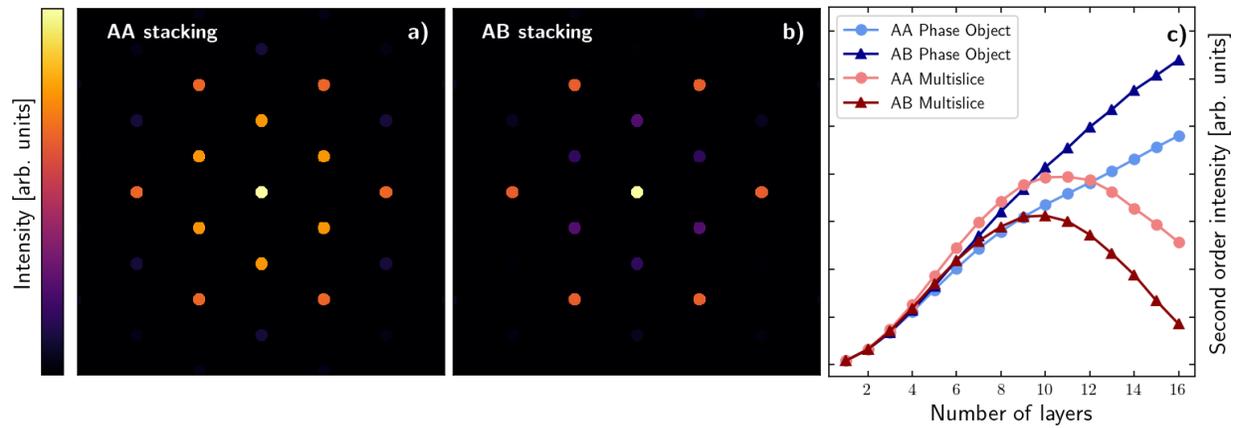

**Fig. S2. (a-b)** Multislice of AA and AB stacked BLG. **(c)** Intensity of second order spots with monotonic increase with N for AA and AB stacked MLG. Both PO and multislice approximation indicate kinematical behavior until N=8. The behavior of the second order spots does not depend on the type of stacking.

## S3. ELLIPSE FITTING OF THE GRAINS

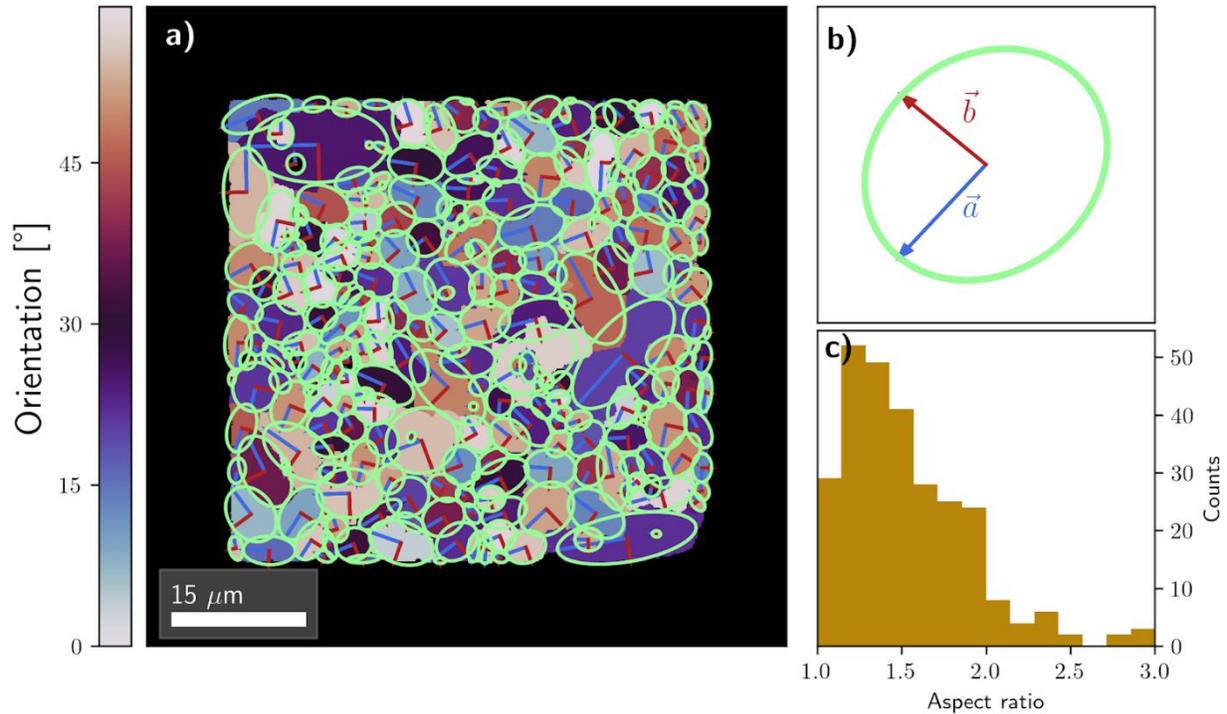

**Fig. S3. (a)** Example of grain shape recognition in the elliptical approximation. The scan of 512x512 probe positions over one square of SLG is the same as Fig.2 in the main text. For every grain, the edges are detected and fitted to an ellipse using least squares fitting. The ratio between the two principal axes is then calculated (blue divided by red). **(b)** Schematic view of one ellipse with the two principal axes are indicated. **(c)** Histogram of aspect ratio of the grains showing that the shape of the grains is not spherical on average.

## S4. FITTING RESULTS GRAIN ORIENTATION AND ASPECT RATIO HISTOGRAMS

The grain orientation histograms (Fig. 4c) are fitted with the sum of two Gaussians with an offset:

$$f(x) = \sum_{i=1}^{2} A_i \cdot e^{-\left(\frac{x-\xi_i}{2\sigma_i}\right)^2} + d$$

where A is the amplitude, $\xi$ the position of the Gaussian, $\sigma$ the variance and d is the offset. The result of the fit of the three different distributions is in Table S1. The amplitude/offset ratio indicates a correlation between the preferred orientations and grain size.

**Table S1.** Quantitative analysis of peaks in the grain size distribution profile (in Fig. 4b)

| Orientation\Size | 0-5 μm | 5-13 μm | 13-192 μm |
|---|---|---|---|
| $A_1$ | 10.71 | 14.11 | 38.29 |
| $A_2$ | 10.00 | 10.69 | 40.17 |
| $\xi_1$ | 20.67 | 21.91 | 23.05 |
| $\xi_2$ | 50.00 | 53.45 | 52.73 |
| $\sigma_1$ | 10.04 | 10.2 | 7.04 |
| $\sigma_2$ | 11.54 | 7.06 | 7.4 |
| d | 19.21 | 20.09 | 7.39 |

The aspect ratio histograms from Fig. 4e are fitted with the skew-normal distribution[13], in order to have a model resembling the data from which we can compare its parameters to give quantitative information on the experimental distributions.

$$g(x) = A \cdot e^{-\left(\frac{x-\xi}{2\sigma}\right)^2} \cdot \left[1 + erf\left(\frac{\alpha(x-\xi)}{\sqrt{2}\sigma}\right)\right]$$

where A is the amplitude, $\xi$ is the location, $\sigma$ is the scale and $\alpha$ is the shape parameter. If $\alpha$ is zero, the normal distribution is retrieved. The erf is the error function.

The average values of the distribution are given by:

$$\langle x \rangle = \xi + \frac{2}{\pi}\sigma\frac{\alpha}{1+\alpha^2}$$

The result is in Table S2, giving the correlation between shape and size.

**Table S2.** Quantitative analysis of ellipse fitting data for grain shape identification (Fig. 4e)

| Aspect Ratio\Size | 0-5 μm | 5-13 μm | 13-192 μm |
|---|---|---|---|
| A | 203.08 | 164.54 | 122.20 |
| $\xi$ | 1.00 | 1.00 | 1.02 |
| $\sigma$ | 0.4 | 0.49 | 0.56 |
| $\alpha$ | 2.62 | 5.99 | 9.25 |
| <x> | 1.31 | 1.40 | 1.46 |

## S5. IDENTIFICATION OF SMALL PARTICLES/SHEETS DISTRIBUTION

4d-STEM in SEM can be applied to cover mm-scale areas and identify small (100 nm-scale) particles/flakes, whereby, in subsequent steps, only the interesting regions are further investigated. This method can link high resolution methods, such as TEM, to lower resolution methods (Raman, P-SHG). Fig. S4 applies this approach to GO. After background subtraction, the flakes are identified. The ML GO region with (blue square in Fig. S4 (d) is discussed in the main text.

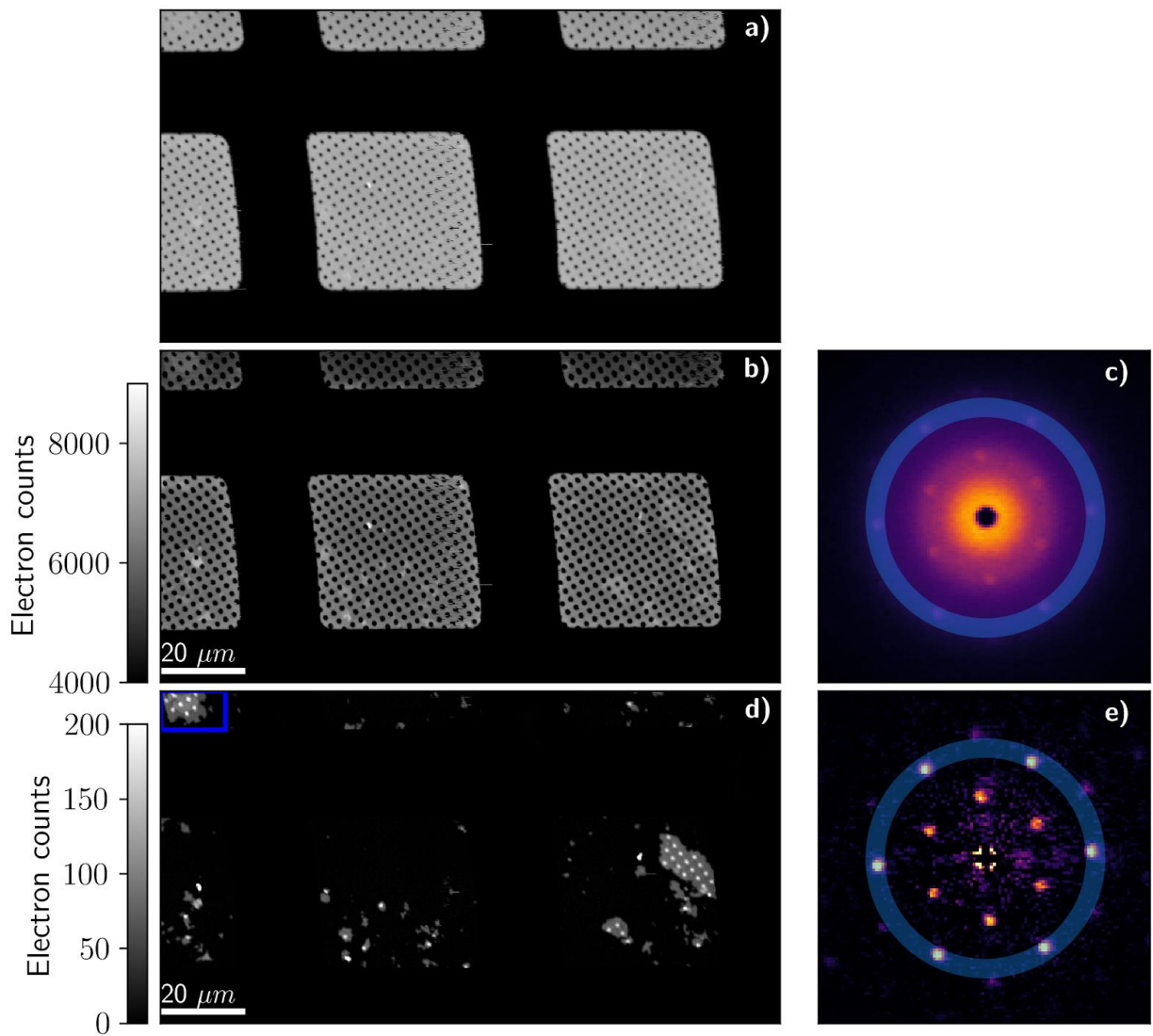

**Fig. S4. (a)** Unprocessed diffraction pattern of 1L-GO. **(b)** VDF signal of the scan. The virtual aperture is placed at the second order spots to increase diffraction contrast. Some increase in the signal is seen where GO is positioned. An overall demise of signal is seen at the top due the instability of the electron gun. The distortions on the image are due to sample drift during the acquisition. **(c)** Unprocessed electron diffraction patterns. **(d)** VDF of subtracted diffraction patterns. The contrast comes from the sparse diffraction intensities. The lighter dots arise from the diffraction pattern inside holes, where the diffraction intensities are larger than on the substrate,

due to additional scattering of the electrons with the substrate. At the top left an overall increase is seen indicating ML-GO. This region is discussed in the main text. **(e)** Diffraction pattern after background subtraction. The background is determined by calculating the azimuthal median for every radius. The center is defined at that of the diffraction pattern leaving only sparse diffraction spots in the signal. This makes it advantageous for crystalline particle/sheets tracking.

## S6. DETERMINATION OF THE ARMCHAIR ORIENTATION IN $MoS_2$ USING P-SHG

SHG imaging is performed in epi-detection using a custom-built laser raster-scanning microscope[14,15], Fig.S5c. A diode-pumped Yb:KGW fs oscillator (1027nm, 90 fs, 76 MHz, Pharos-SP, Light Conversion) is inserted into a modified, inverted microscope (Zeiss Axio Observer Z1), after passing through 2Ag coated galvanometric mirrors (6215H, Cambridge Technology). A motorized rotation stage (M-060.DG, Physik Instrumente) with a zero order $\lambda/2$ wave plate (QWPO-1030-10-2, CVI-Lase) is used to rotate the direction of the excitation linear polarization. The beam is then reflected on a dichroic mirror at 45° (DMSP805R, ThorLabs), placed at the turret box of the microscope, just before the objective (Plan Apochromat 40x 1.3NA, Zeiss). The SHG signal is collected from the same objective used for excitation and guided onto a photomultiplier tube (PMT) detector (H9305-04, Hamamatsu). In front of the PMT, a home-built mount holds a bandpass filter (FF01-514/3-25, Semrock) and a short pass filter (FF01-680/SP-25, Semrock) for SHG imaging. After the filters, a film polarizer (LPVIS100- MP, ThorLabs) is inserted in front of the PMT to measure the anisotropy of the SHG signals due to the rotation of the excitation linear polarization. Coordination of PMT recordings with galvo-mirrors movements and with all the motors, as well as the image formation, is done in LabView

The SHG from a 1L-MoS$_2$ is described by its corresponding $\chi^{(2)}$[14]. For 1L-MoS$_2$ (D$_{3h}$) the $\chi^{(2)}$ tensor elements that contribute to SHG are[15]:

$$\chi^{(2)}_{xxx} = -\chi^{(2)}_{xyy} = -\chi^{(2)}_{yyx} = \chi^{(2)}_{yxy},$$

where, x,y,z denote the 1L-MoS$_2$ crystal coordinates system, x being along MoS$_2$ armchair direction (Fig. S5a). The laser propagates in the Z-axis and the x armchair direction of 1L-MoS$_2$ is at an angle $\theta$ with the X-axis (Fig. S5a). The SHG signal when rotating the excitation linear polarization and detecting the SHG parallel with the X-axis, is given by[15]:

$$I_s^{2\omega} = [A\ \cos(3\theta - 2\varphi)]^2,$$

where $A = E_0^2 \varepsilon_0 \chi^{(2)}_{xxx}$, $\varepsilon_0$ is the vacuum permittivity and $E_0$ is the amplitude of the excitation field. The graphical representation of the equation above and the corresponding visualization in a polar diagram (Fig. S5b) demonstrate a fourfold symmetry of the P-SHG intensity modulation that rotates for different $\theta$. Thus, each $\theta$ corresponds to a characteristic fourfold symmetric ("four leave rose like") polar diagram. We calculate $\theta$ by fitting to the experimental parallel to the X-axis component of the SHG signal for different linear excitation directions.

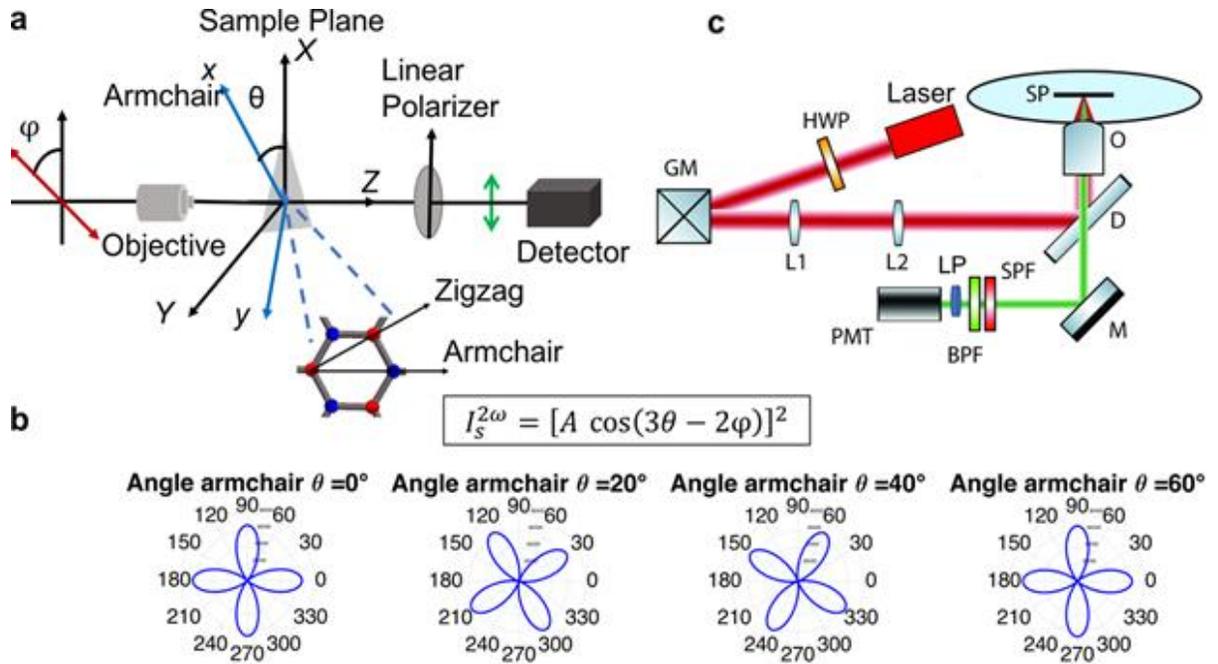

**Fig. S5. (a)** Coordinates system for SHG measurements. X and Y refer to the lab coordinates, x and y are the crystal coordinates, θ is the angle between the X-axis and the armchair direction and needs to be determined. φ is the angle between incident wave polarization and the lab X-axis, which is what we control in the experiment. The crystal structure and symmetry axis of MoS$_2$ are shown (top view) magnified. **(b)** Polar diagrams of the parallel component of the SHG intensity as a function of polarization angle for θ=0°, 20°, 40° and 60°. **(c)** Schematic P-SHG setup. HWP, half-wave plate; GM, galvanometric mirrors; L1,2, achromatic lenses; D, dichroic; O, objective; SP, sample plane; M, mirror; SPF, short-pass filter; BPF, bandpass filter; LP, linear polarizer; PMT, photomultiplier tube. The excitation linear polarization starts horizontal in SP and is rotated 0°–360° with a step size of 2°.